\documentclass[twocolumn]{svjour3} 

\pdfoutput=1
\usepackage{graphicx}
\usepackage[caption=false]{subfig}
\usepackage{color}
\usepackage{float}

\begin{document}

\title{
The leverage effect and other stylized facts displayed by Bitcoin returns
}
\author{F.N.M. de Sousa Filho \and J.N. Silva \and M.A. Bertella \and E. Brigatti}

\institute{F.N.M. de Sousa Filho \at Instituto de Matem\'atica, Universidade Federal do Rio de Janeiro,
Av. Athos da Silveira Ramos, 149,
Cidade Universit\'aria, 21941-909, Rio de Janeiro, RJ, Brazil\\
\and
J.N. Silva \at Deparment of Economics, S\~ao Paulo State University (UNESP),
14800-901 Araraquara, SP, Brazil\\
\and
M.A. Bertella \at Deparment of Economics, S\~ao Paulo State University (UNESP),
14800-901 Araraquara, SP, Brazil\\
\and
E. Brigatti
\at Instituto de F\'{\i}sica, Universidade Federal do Rio de Janeiro,
Av. Athos da Silveira Ramos, 149,
Cidade Universit\'aria, 21941-972, Rio de Janeiro, RJ, Brazil\\
 \email{edgardo@if.ufrj.br} } 


\maketitle

\begin{abstract}
In this paper, we explore some stylized facts of the Bitcoin market  using the BTC-USD exchange rate time series of 
historical intraday data from 2013 to 2020.
Bitcoin presents some very peculiar idiosyncrasies, like the absence of macroeconomic fundamentals or connections with underlying assets or benchmarks, an 
asymmetry between  demand and supply and the presence of inefficiency in the form of 
strong arbitrage opportunity. Nevertheless, all these elements seem to be marginal in the definition of the structural statistical properties of this virtual financial asset, which result to be analogous to general individual stocks or indices.
In contrast, we find some clear differences, compared to fiat money exchange rates time series,
in the values of the linear autocorrelation and, more surprisingly, in the presence of the leverage effect.
We also explore the dynamics of correlations, monitoring the shifts in the evolution of the Bitcoin market. This analysis is able to distinguish between two different regimes:
a stochastic process with weaker memory signatures and closer to Gaussianity
between the Mt. Gox incident and the late 2015, and a dynamics with relevant correlations and strong deviations from Gaussianity before and after this interval.

\keywords{ Fluctuation phenomena \and Random processes \and Noise \and Brownian motion}


\end{abstract}

\section{Introduction}

A cryptocurrency is a digital currency that can theoretically perform all the three functions of money, namely, medium of exchange, unit of account and store of value. 
Until now, the spread of this  type of money is still in its infancy, despite its considerable numbers: it is estimated that today 
there are about 5,000 cryptocurrencies with a market capitalization of the order of US\$240 billions \cite{CoinMarketCap2020}.
The first and most important cryptocurrency to date is Bitcoin. It was developed in 2008 and diffused in a paper assigned using the pseudonym of Satoshi Nakamoto \cite{Nakamoto2008}. Bitcoin is an open source, peer-to-peer currency, and does not depend on monetary or governmental authority. Its invention is revolutionary given that there is no need to have an agent like a bank to effect the transaction, only the two parties involved: the payer and the receiver of the debt. All transactions carried out are stored in a public register called blockchain so that future transactions are unable to use the previously spent Bitcoins. Note that transactions on the Bitcoin network are not made in another fiat currency but they are made in Bitcoins. Thus, in addition to being a completely decentralized network, it is a virtual currency, where its value is defined in a market, such as the dollar, euro, swiss franc, etc. 
As its use has been quite restricted, many economists still do not consider it as currency, as it lacks at least one of its typical functions. In fact, Baur {\it et al.} \cite{BAUR2018} notes that, for the period from 2011 to 2013, most Bitcoins were used as a portfolio asset (value store) and not as a currency (medium of exchange).
In fact, most of the demand as a store of value has been directed towards Bitcoin due to the credibility and predictability of its supply and resilience demonstrated during its short existence. The Bitcoin
supply will increase around 1\% per year for the next 25 years \cite{Ammous2018}, whereas the main currencies, like US dollar and Euro, will rise much more than that, specially if the quantitative easing remains as one of the main tools of Central Banks monetary policy. Although Bitcoin has not fulfilled all the three functions of money, it is possible that Bitcoin will continue to draw more interest as a store of value and to play a broader position as a medium of exchange. However, we can not say the same for other cryptocurrencies, which, as a store of value or a unit of account, do not seem to offer any advantages and thus are unlikely to attract interest as a medium of exchange. For instance, Ether, the second largest virtual currency by total market cap, has an annual growth rate much higher than Bitcoin (an estimated average of around 5\% between 2020 and 2049) and the Ethereum Foundation (the entity behind this currency) has a discretionary power to change the issuance of Ether \cite{Ammous2018}. A presence of a clear and credible commitment to a monetary issuance policy similar to that of Bitcoin lacks to other cryptocurrencies (Ether, Dogecoin, and so forth) and this seems to be one of the main reasons of Bitcoin success. The only digital currency that can credibly show a degree of severe adherence to its issuance timetable is Bitcoin, which gives assured protection to future holders. 
In sum, the inflexible supply of digital currencies and their volatile demand make them unreliable for use as unit of account. Only Bitcoin may be able to act as a store of value due to its strict dedication to low supply expansion, convincingly supported by the distributed protocol of the network and a reliable proof of the lack of any entity capable of modifying the supply timeline. Centrally managed governance over other digital currencies and the application of tokens with unique purposes render them unable to perform monetary functions.

Empirical studies \cite{BAUR2018,LIU2018,bariviera,URQUHART2019}  
show that Bitcoin and other cryptocurrencies generally present time series characterized by the following simple descriptive statistics: high returns, high volatility, 
important skewness and high kurtosis. Moreover, some return autocorrelations and a changing behavior of the Hurst exponent has been detected.

The study of autocorrelations and its connection with the presence of short or long range memory is a particularly interesting topic because of its relation with the efficient market hypothesis (EMH). This idea implies determining whether asset prices fully reflect all available information \cite{fama}. If so, it means that any new information is revealed entirely in its price. The question here is to know what kind of information might be taken into consideration for a market to be considered efficient. The most widely used version of EMH is the semi-strong: prices accurately reflect all publicly known information. Under this version, there is no cheap or expensive asset: the current price is always its best estimate. This definition implies that the knowledge of past prices is useless to predict future prices, which leads us to the weak version of EMH: historical prices have no relevance to predict prices. From a statistical viewpoint, this means that asset prices or returns cannot present long range memories, since they would allow a riskless profitable trading strategy. Finally, there is the strong version of EMH: the prices accurately reflect all the information (public and private). While the first two forms of market efficiency, weak and semi-strong, have numerous advocates, there is a general perception that the strong version is difficult to empirically validate.


Since the 1980s, EMH has been questioned and the analysis of the presence of long memory in financial time series has become an important topic, with papers presenting some empirical evidences of long memory \cite{noFama}, and others challenging it \cite{okFama}.  
Several studies have been done to draw some conclusion about the efficiency of Bitcoin, however without a definitive one 
(see \cite{bariviera,Bartos2015,Urquhart2016,Nadarajah2017,Dimitrova2019,Nan2019} among others). 
Bartos \cite{Bartos2015} analyses the effects of public announcements on Bitcoin price concluding for a positive answer. Therefore, its price seems to follow the EMH. Urquhart \cite{Urquhart2016}, through several tests, finds that Bitcoin market is inefficient, but when he splits into two subsample periods, he finds that the Bitcoin may be in a process towards efficiency. Nadarajah and Chu \cite{Nadarajah2017} show, through eight different tests, that Bitcoin satisfies the EMH. Dimitrova {\it et al.} \cite{Dimitrova2019} conclude that, although the self-similarity exponent of the BTC-USD price series is different than 0.5, this result is not due to the presence of significant memory but to its underlying distribution. Bariviera {\it et al.} \cite{bariviera} analyze the behavior of long memory of returns from 2011 until 2017, using the Hurst's exponent. They show a persistent behavior from 2011 until 2014, whereas the series is more informational efficient since 2014. Finally, Nan and Kaizoji \cite{Nan2019} study the Bitcoin market efficiency in terms of the Bitcoin exchange rate and conclude that the weak and semi-strong form of market efficiency of the USD/EUR Bitcoin exchange rate holds in the long run concerning the spot, futures, and forward FX markets.

These analysis are of particular relevance since the common sense speaks about inefficiency of the cryptocurrency market. Arbitrage opportunity are evident, as the price may present an enormous spread among the different platforms used for buying and selling the cryptocurrency. Moreover, Bitcoin price is not driven by macro-financial indicators, a fact that can generate more sensitivity on information flows in market affecting the supply and demand interaction. For these reasons, a natural question that emerges is whether these inefficiencies can be tracked in the Bitcoin historical time series, either in the form of a clear long memory, 
or in more subtle temporal structures present in the data.

With this aim, our work will try to systematically characterize the empirical properties of the returns of the Bitcoin time series, highlighting the most relevant stylized facts \cite{cont,cont2} present in these data. Among them, we will describe the heavy tails of the distribution of returns, and the autocorrelations of some nonlinear functions of returns. 
These features have been  studied in recent works with similar or different approaches.
A comparison with these previous results will be unfolded in the 
Sections 4 and 5 of the paper. 
The important novelty of our work is the clear characterization 
of the presence of  the leverage effect in Bitcoin, and, more in general, in an exchange rate return, by
measuring the realized volatilities in intraday data.
A comparison of all the examined stylized facts 
with well known features already detected in the time series of the exchange rate of fiat currencies, which are among the most liquid assets in the world, could shed light on the real impact that inefficiencies can have on the statistical properties of the temporal dynamics of a general financial asset. 

This approach will be used also for characterizing the time varying behavior of the evolution of Bitcoin returns (Section 4.2). In this way we will explore the possibility of using some of these statistical features as empirical indicators or signals for monitoring the shifts in the evolution of the Bitcoin market. 
This analysis will show some interesting 
aspects related to how
important events and publicly announced information can affect the prices of cryptocurrencies. 

\section{Data} 

\begin{figure}[!h]  
\begin{center}
\includegraphics[width=0.45\textwidth, angle=0]{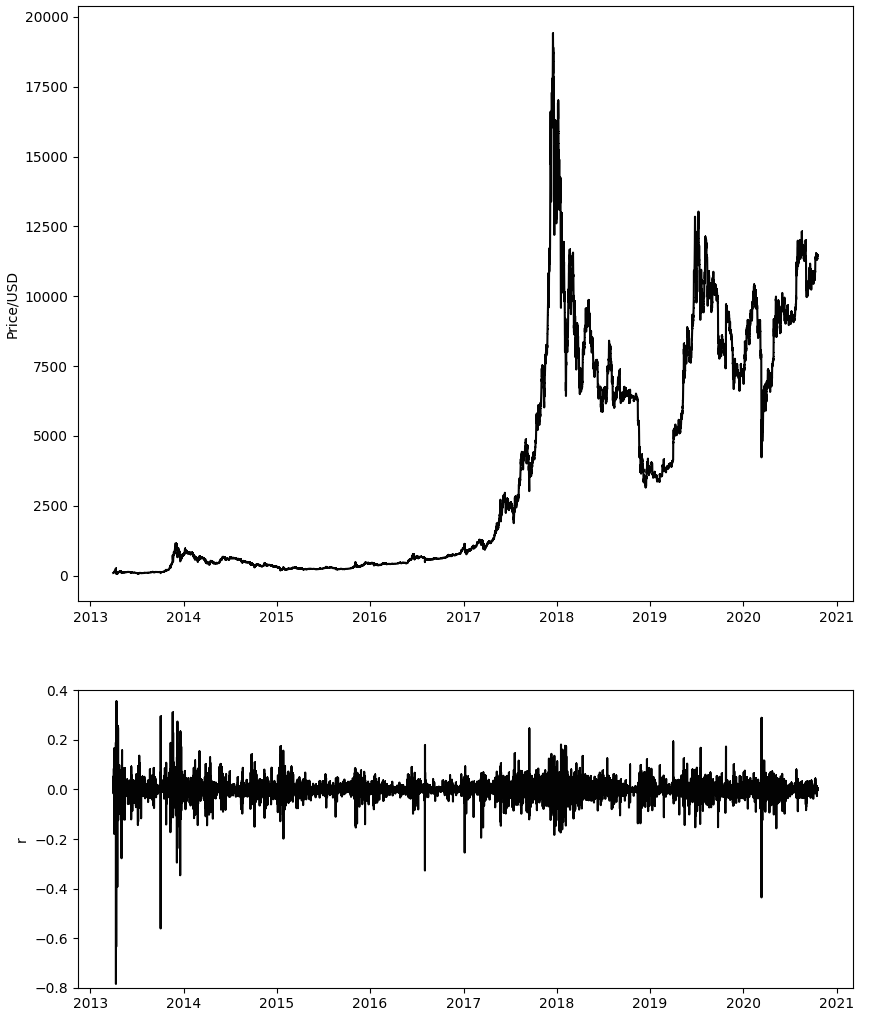}
\end{center}
\caption{\small {Price and return (on the bottom) of Bitcoin 
expressed in Dollars from 31/3/2013 to 19/10/2020 (time lag of 5 hours).
}
}
\label{fig_Data}
\end{figure}

We examine the time series of the price of Bitcoin, expressed in US dollars, 
from 31/3/2013 to 19/10/2020, with a sampling time interval of 5 hours,
which corresponds to a series of 13246 elements (see Fig. \ref{fig_Data}).
We leave out earlier periods due to low market liquidity.
Data are extracted from  the site Bitcoincharts \cite{BitChar} and are representing the
Bitcoin price 
from the Bitstamp exchange market platform.
Note that Bitcoin price, in the considered period, can depend strongly
on the platform used for trading.
At high frequencies, data directly downloaded from Bitcoincharts do not present regular
time intervals of sampling. For this reason, we realize a resampling of the series with intervals of 5 hours. This resampled 
time series presents a good statistical quality.\\

\section{Stylized facts: definitions and methods}

Empirical properties of traditional financial assets 
have been characterized by a set of regularities and general tendencies 
known as ``stylized facts" \cite{cont,cont2}.
This term was introduced by the economist N. Kaldor,
who pointed out the importance of highlighting 
broad and robust tendencies of statistical dataset, by ignoring individual details \cite{Kaldor}.
In finance, stylized   facts are   usually formulated  in  terms  of 
statistical  properties of the time series of asset returns which present a character of universality,  being 
present in different scenarios and 
common to a wide variety of markets, instruments and time periods.
Their identification could be helpful for
validating the approaches used to study financial data and
for establishing empirical based  models
in order to produce reliable forecasts.   
Among the different  stylized facts recently recorded, 
in this article  we will consider a set of facts 
presents in the time series of the returns of a single asset and now widely accepted. 
In table \ref{table0} we present a synthetic representation of
these stylized facts and in the following 
we  explain in details the methods used for their analysis.\\

\begin{table}
    \centering
{\small
    \begin{tabular}{|c|c|c|}  
\hline
{\bf Stylized Fact} & {\bf Definiton} & {\bf Implications} \\
\hline
\hline
{\tiny \bf Heavy tails} & {\tiny  Distribution of returns  } & {\tiny Finite second moment }
 \\
 &   {\tiny displays a power-law tail } &  {\tiny (fundamental for risk } \\
 & { \tiny (tail index $\sim 3$).  Increasing  }  & {\tiny management and}\\
 &  {\tiny  $\Delta t $, the distribution }    &  {\tiny portfolio optimization)}\\
 & {\tiny approaches  a Normal one. } & \\
\hline
{\tiny \bf Absence of linear}  &  {\tiny Autocorrelations of returns  } &  {\tiny Possible evidence} \\
{\tiny \bf autocorrelations}  & {\tiny are insignificant, except for } & 
{\tiny for the efficient market}\\
    & {\tiny very  small intraday scale} &  {\tiny hypothesis} \\
\hline
 {\tiny \bf Persistence in} & {\tiny Autocorrelation of absolute  } & {\tiny  Long-range} 
 \\
{\tiny \bf absolute returns}   & {\tiny returns decays as a power-law} & {\tiny dependence } \\
\hline
{\tiny \bf Volatility }  & {\tiny  Different measures of  }  
&  {\tiny High-volatility tends } \\
 {\tiny \bf clustering}  & {\tiny volatility display persistence} & {\tiny to cluster in time}\\
\hline
{\tiny \bf Leverage effect}  & {\tiny Volatility are negatively } &  {\tiny Volatility increases} \\
 & {\tiny correlated with returns} &  {\tiny with negative shocks} \\
\hline
\hline
\end{tabular}
}
\caption{ The set of stylized facts considered in this article.
}
\label{table0}
\end{table}

As usual, the characterization of stylized facts is realized using the return of the price series $r(t)$, 
defined as:
\begin{equation}
r(t)=\ln(P(t))-ln(P(t-\Delta t))
\end{equation}
where $P(t)$ is the price at time $t$ and $\Delta t$
is the considered sampling time interval (for our series, $\Delta t$ corresponds to 5 hours).

As a first step, we study the distribution of returns,
with the aim of characterizing the expected presence
of heavy tails \cite{cont,cont2}.
The non-Gaussian shape of the distribution of price changes 
is a well known character when sufficient small sampling time intervals are considered.
It can be quantified  looking at the  kurtosis of the distribution.
More interesting is the description of the 
behavior of the  tail of the distribution.
Based on some general empirical results 
for foreign exchange markets and  for stock markets \cite{guillame,gabaix}, 
the tail can be characterized using a power-law distribution.
For a better comparison with similar works present in the literature,
the distribution of the price change is defined using 
a normalized return: $r_{N}(t)=\frac{r(t)- \langle r \rangle}{\sigma}$,
where $\langle r \rangle$ is the mean value of the returns and $\sigma$
the standard deviation estimated over the whole time series.
We estimate the power-law exponent $k$ (tail index) of the cumulated distribution
using the Maximum Likelihood estimate (see, for example \cite{power1}).
As usual, we use only the data larger than
the smallest value for which the power-law behaviour holds ($|r^{min}_{N}|$).
This lower bound on the power-law behaviour  is estimated 
following an approach proposed by Clauset {\it et al.} \cite{power3}:
we select the value that makes the probability distribution of the measured 
data and the best-fit power-law model as similar as possible above $|r^{min}_{N}|$.
\\

We continue our analysis studying the dependence and
memory properties of the time series.
We analyse the autocorrelation function:
\begin{equation}
A(\tau)=corr(r(t+\tau),r(t))
\label{eq1}
\end{equation}
where $corr()$ is the Pearson's correlation between the two variables.
In liquid markets this linear autocorrelation function is 
expected to reach zero in a few minutes, with really faster decays in the most liquid ones, like
the foreign exchange markets \cite{corr2,corr}.
This fact has been often cited for supporting the efficient market hypothesis \cite{fama}.
In addition, we tested the autocorrelation present in our time series with 
a non-parametric measure, by using the Spearman's rank-order correlation \cite{spearman}.
The Spearman's correlation between two variables is equal to the Pearson's correlation between the rank values $R$ of those two variables. In our case, the Spearman's autocorrelation function corresponds to: $A_{S}(\tau)=corr(R_{r(t+\tau)},R_{r(t)})$.
More details can be found in \cite{Lehmann}.\\

For assuring the independence of the elements of the time series
not only the linear autocorrelation, but any nonlinear function of returns should
present no autocorrelation.
This property is not satisfied by financial time series.
Absolute and squared returns show important  positive autocorrelations,
a fact generated by the tendency of large price variations to be 
followed by large price variations (volatility clustering).
For this reason we will analyse 
the autocorrelation functions of powers of absolute returns:

\begin{equation}
A_{\alpha}(\tau)=corr(|r(t+\tau)|^{\alpha},|r(t)|^{\alpha})
\end{equation}
considering the case with $\alpha=1$ and $\alpha=2$.
The first one generally presents the highest correlations,
and the second one is commonly used for measuring volatility clustering.

The decay of these autocorrelation functions 
are usually well described by a power law (see \cite{power}):
$A_{\alpha}(\tau)\propto \tau^{-\beta}$. 
For $\alpha=1$ and $\alpha=2$ the coefficient 
$\beta \in [0.2, 0.4] $, as reported in \cite{power2}.\\

Finally, we measure the correlation of returns with subsequent squared returns:

\begin{equation}
L(\tau)=corr(|r(t+\tau)|^2,r(t)).
\label{eq:leverage}
\end{equation}

It was shown empirically that this measure generally starts from a negative value for $\tau\approx0$ and it grows towards zero \cite{corr2,leverage0}, suggesting that negative returns generate a rise in volatility. 
This negative correlation of volatility with returns is usually named ``leverage effect".

We introduce a simple way for characterizing the presence of the leverage effect 
in our time series, defining a critical correlation time $\tau_0$, which estimates the temporal 
scale over which the leverage effect (anticorrelation) is relevant.
This critical time corresponds to the smallest value of $\tau$ where $L(\tau)$ crosses zero. 
We can easily estimate $\tau_0$ calculating the
cumulative leverage $L_c(\tau)=\sum_{i=0}^{\tau}L(i)$.
Note that, if the time series presents the leverage effect, $L(0)$ is negative
and $L_c(\tau)$ is a convex function with a well defined minimum.
In this case $\tau_0$ is the $\tau$ value where $L_c(\tau)$ reaches the minimum \cite{io}.
In fact, at this lag time the leverage vanishes and 
the volatility no more exercises a negative  influence on the return.
If $L(0)$ is not clearly negative and $L_c(\tau)$ does not 
present a convex shape the leverage effect is not present.

\section{Results}

\subsection{General results}

An inspection of simple descriptive statistics of 
the considered time series gives a Standard Deviation value of 0.032,
a Skewness of -2.93 and a Kurtosis of 72.64.
These results are consistent with previous results 
appeared in the literature \cite{bariviera} which, in relation to other traditional currencies, 
pointed out an extreme high value of the volatility, with a difference of one order of 
magnitude and
comparable values for higher moments.



\begin{figure}[!h]
\begin{center}
\includegraphics[width=0.45\textwidth, angle=0]{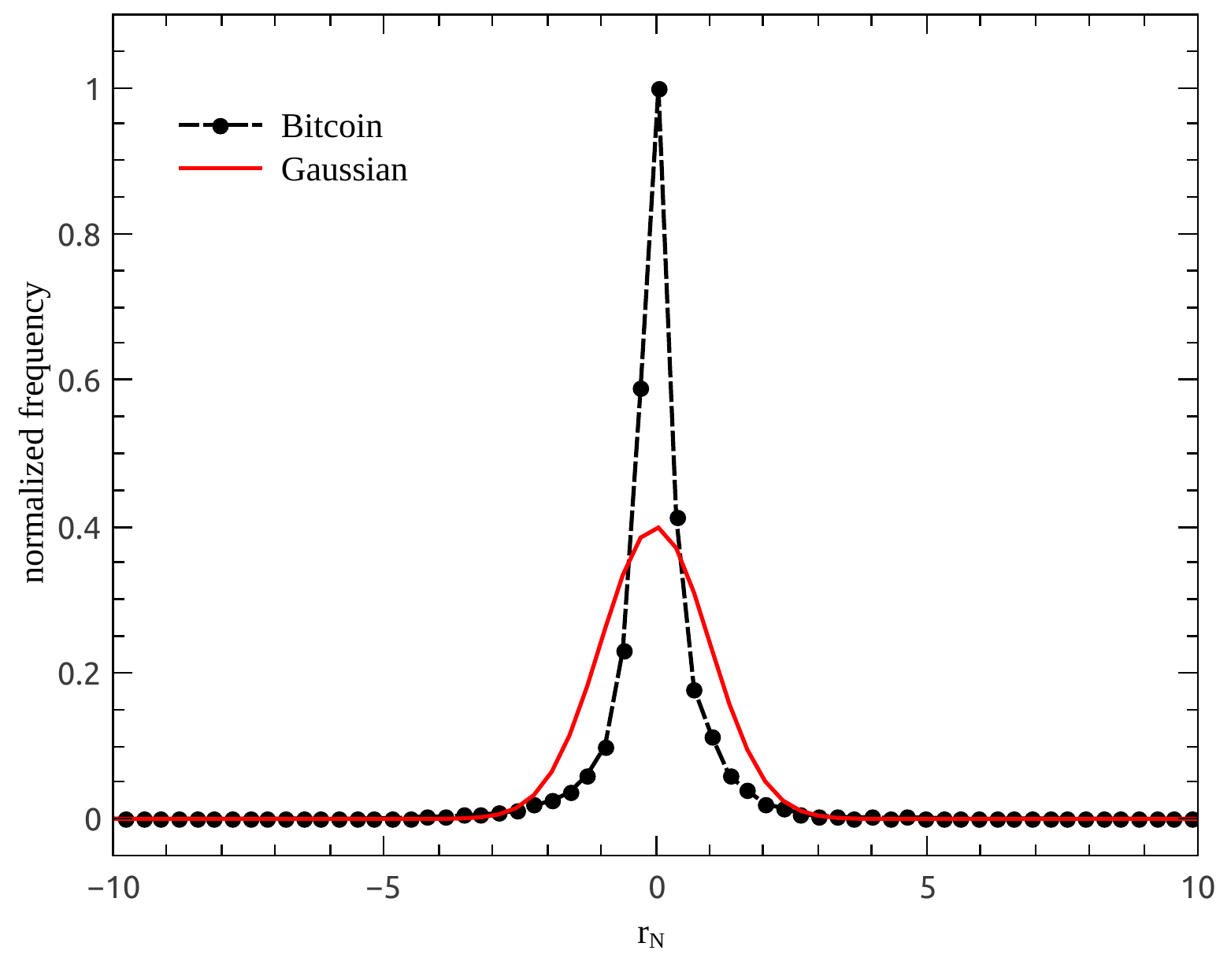}
\includegraphics[width=0.45\textwidth, angle=0]{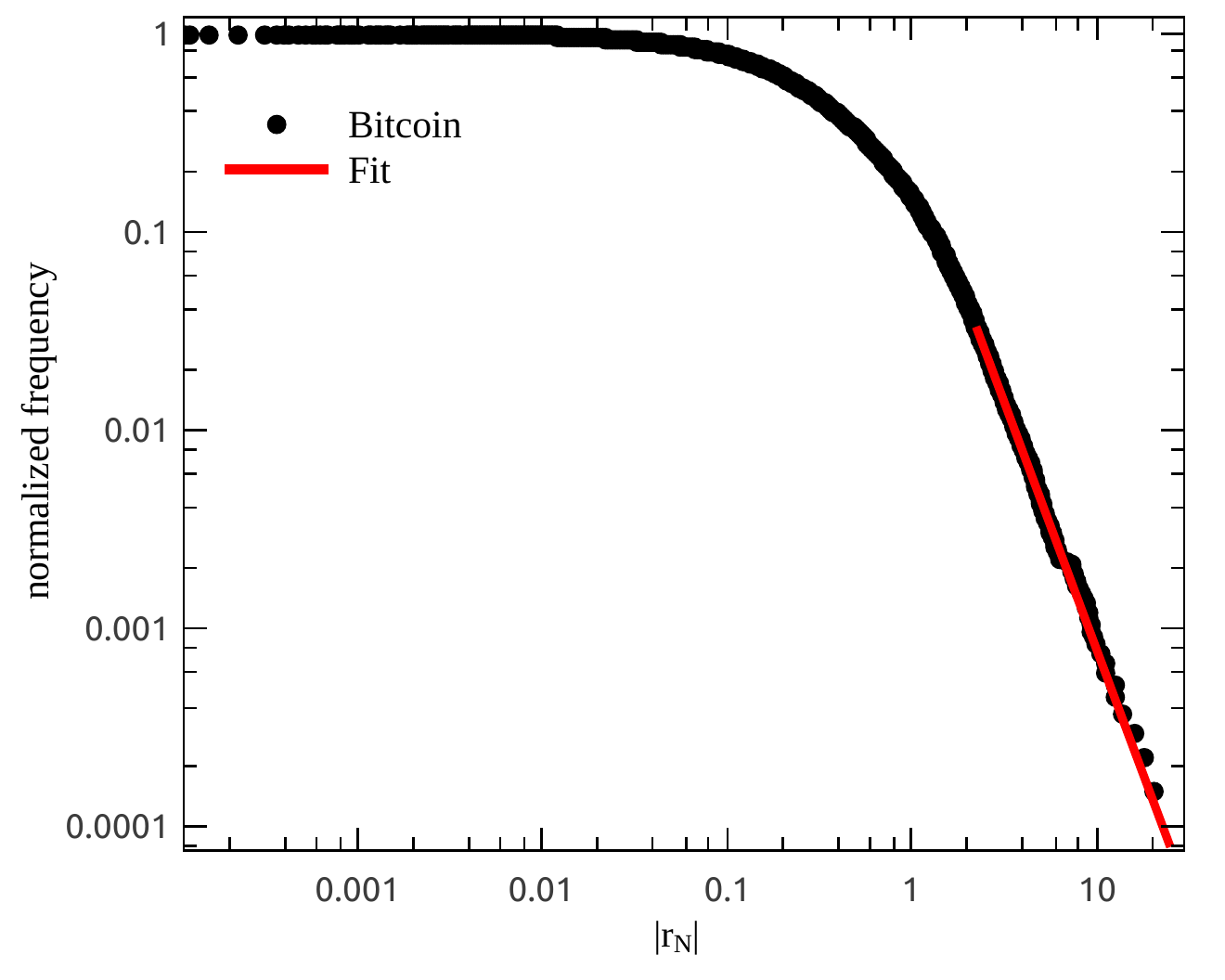}
\end{center}
\caption{\small {{\it 
Top:} Normalized returns distribution of Bitcoin.
A Gaussian fitting is represented by the red continuous line.
{\it 
Bottom:} Log-log plot of the cumulated 
distribution of the absolute value of the normalized returns.
The red continuous line is the power-law fitting of the higher values of the distribution ($|r^{min}_{N}|=2.28$).}
}
\label{fig_distr}
\end{figure}

In Fig. \ref{fig_distr} we present the Bitcoin distribution of the normalized returns $r_N$, 
with its obvious non-Gaussian shape.
The estimation of the tail index of the cumulated distribution of the absolute value of the normalized returns gives a value of $k=3.49\pm0.02$, 
in good correspondence with the ones measured for 
the spot intra-daily foreign exchange markets \cite{guillame}
and  for stock markets \cite{gabaix}, where
the power law estimation 
presented exponents close to 3.


\begin{figure}[!h]
\begin{center}
\includegraphics[width=6.5cm,height=4.5cm,angle=0]{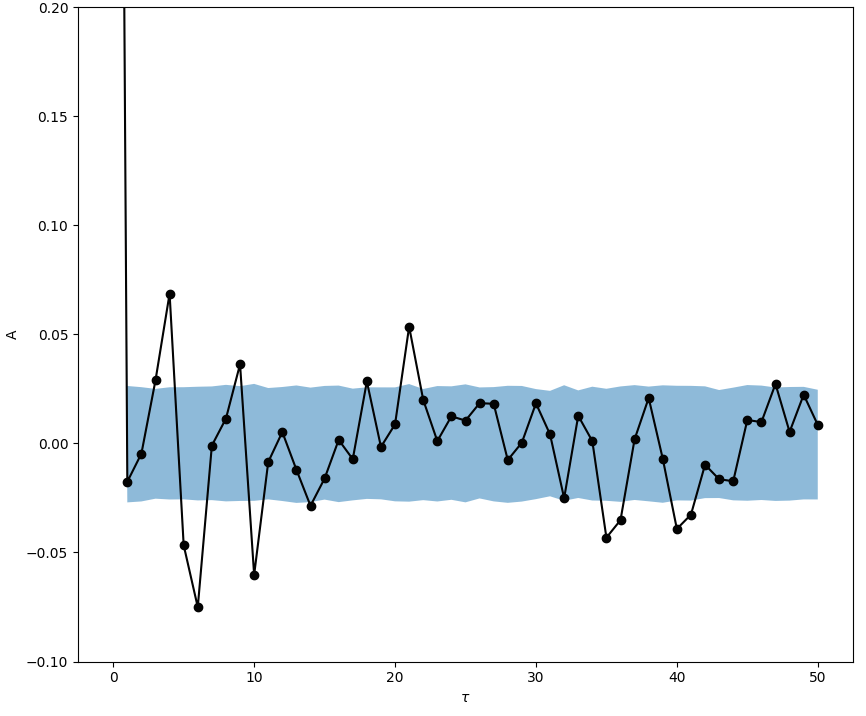}
\includegraphics[width=6.5cm,height=4.5cm,angle=0]{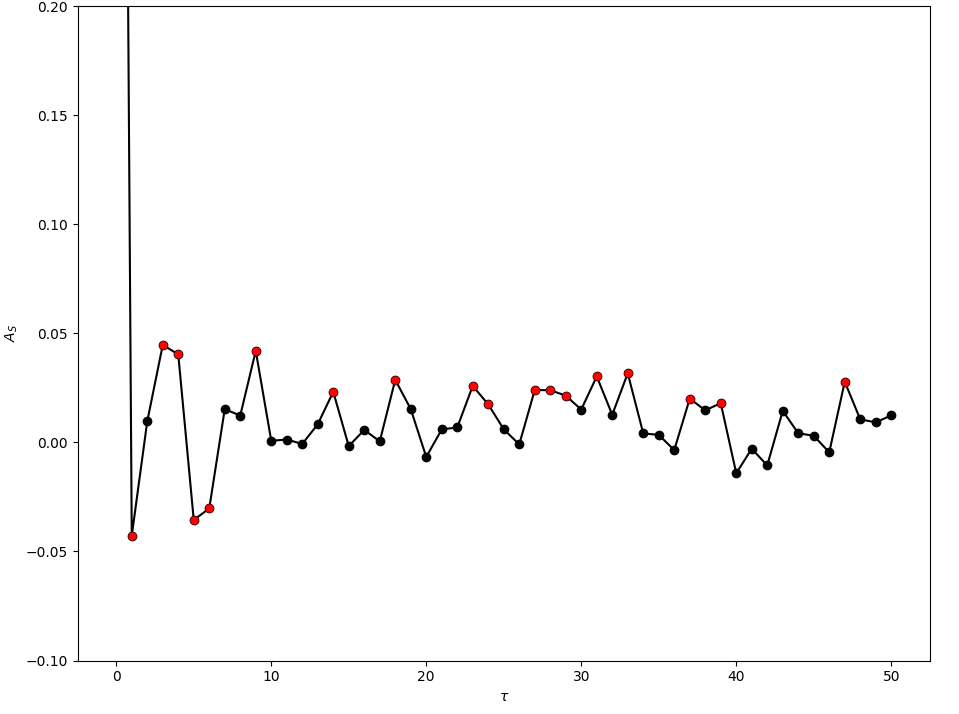}
\end{center}
\caption{\small {Autocorrelation for returns of Bitcoin. 
{\it 
Top}: Pearson autocorrelation; the light blue band represents the 1-/99-percentiles of the distribution of the linear autocorrelation of 
randomized data estimated over 1000 realizations. 
{\it 
Bottom}: Spearman autocorrelation. Red points correspond to autocorrelation values with 
a $p$-value less than 0.05 (the null hypothesis  corresponds to absence of autocorrelation).
}
}
\label{fig_corr}
\end{figure}

In Fig. \ref{fig_corr} we displayed the results relative to the values of
the linear autocorrelations for our dataset.
For the Pearson's autocorrelation, as the empirical observations are clearly not Gaussian,
the significance levels are estimated
comparing the correlation function to the 1-/99-percentiles 
of a generated ensemble of randomized data.
In this case, the results show that the linear autocorrelations are close to be 
negligible.
Anyway, we can detect some values slightly
above the significance level.
Also the Spearman's autocorrelations 
present statistical significant values, in particular for $\tau\le6$.
Spearman autocorrelations are smaller than
the Pearson ones. This is probably due to important observations 
present in the tails of the distributions which positively impact
the Pearson's correlation.
Our general findings in the measure of the autocorrelation function are consonant with some previous 
studies \cite{bariviera},
in particular with the work of Urquhart \cite{Urquhart2016} which showed
some antipersistent behaviors dependent on the sampling period.

The analysis of the behaviour of the absolute and squared returns 
for the Bitcoin time series
follows the behaviour already recorded for other financial
assets.
In fact, it presents important positive autocorrelations 
which decay with a power-law.
The exponents of the best fitted functions are $0.31\pm0.01$
for $\alpha=1$ and $0.72\pm0.09$ for $\alpha=2$.
This second value is larger  than expected, which corresponds 
to a faster decay  in relation to the behaviour reported in \cite{power2}.


\begin{figure}[!h]
\begin{center}
\includegraphics[width=0.5\textwidth, angle=0]{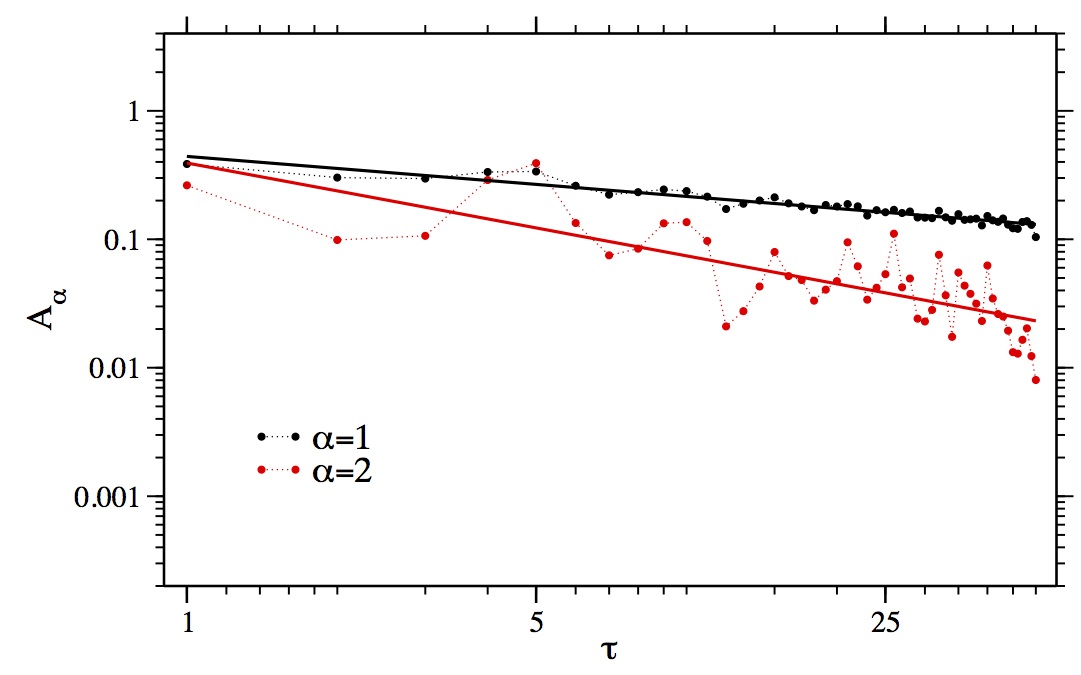}
\end{center}
\caption{\small {Log-log plot of the autocorrelation functions of powers of the returns $A_{\alpha}(\tau)$ for $\alpha=1$ and $\alpha=2$. The continuous lines are the best fitted power laws.
The value of autocorrelations is quite significant; for example $A_1(1)=0.39$ and $A_2(1)=0.26$.
}
}
\label{fig_nonlinear}
\end{figure}

\begin{figure}[!h]
\begin{center}
\includegraphics[width=0.45\textwidth, angle=0]{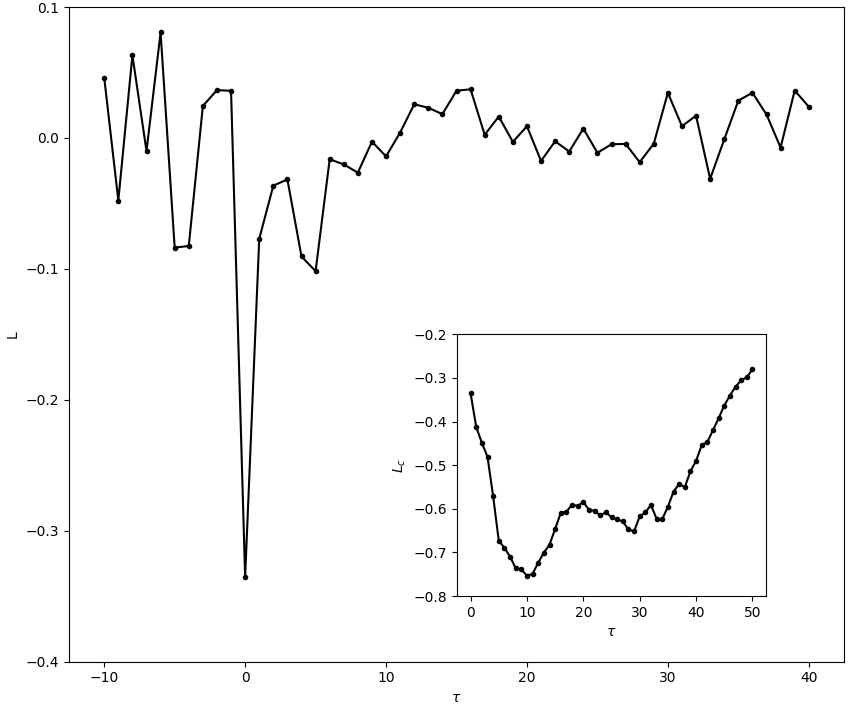}
\end{center}
\caption{\small {  
The leverage effect for the Bitcoin time series. 
A negligible correlation exists for $\tau<0$, which becomes a relevant 
anticorrelation for $\tau>0$. In the inset the cumulated function $L_c(\tau)$, which
is a convex function. 
In this case, the time series presents the leverage effect
and $L_c(\tau)$ reaches the minimum for $\tau_0=10$. 
}
}
\label{fig_leverage}
\end{figure}

The measurement of the 
volatility-return correlation 
shows that the values are not significantly different from zero for $\tau<0$,
whereas they are significant and negative for $\tau>0$.
This behaviour is consistent with the so-called leverage effect (see Fig. \ref{fig_leverage}).
It follows that the cumulated function $L_c(\tau)$ is a clear 
convex function, and $L_c(\tau)$ reaches the minimum for $\tau_0=10$,
which  delineates a 
correlation with a decay time slightly longer than 2 days.
Leverage effect has been previously detected 
for individual stocks and stock indices \cite{leverage0,leverage},
with a typical decay time of about 50 days for 
stocks and 10 days for indices.
It is interesting to note that our result is particularly visible. 
In fact, it does not need to rely on an averaging procedure, as for individual stocks, and the negative correlation is larger than the one usually seen in stock indices.
In general, it is common to find empirical studies
which present
evidence of leverage in stock markets.
In contrast, for foreign exchange returns
the typical assumption is the absence of such 
phenomenon \cite{Hull}, as suggested by 
a large amount of empirical evidences \cite{NoLeverage}.



\subsection{Running windows}

In order to describe the dynamics of correlations all along 
the evolution of the Bitcoin time series, we partition our dataset using sliding 
windows.
We consider windows of 1600 data points, which correspond 
to less than one year. The starting points of the
windows are considered as the time stamp.

\begin{figure}[!h]
\begin{center}
\includegraphics[width=.5\textwidth, angle=0]{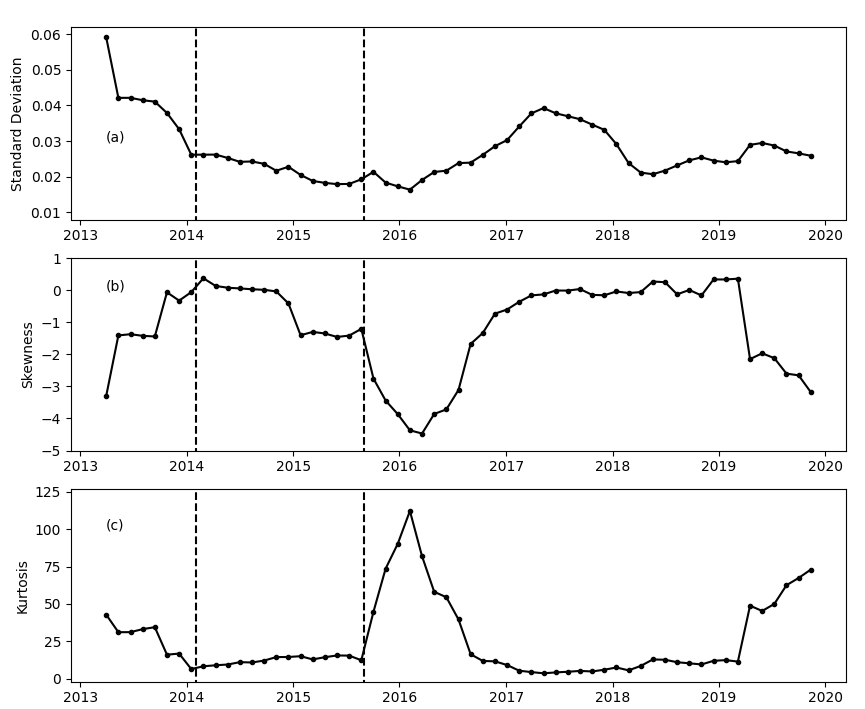}
\includegraphics[width=.5\textwidth, angle=0]{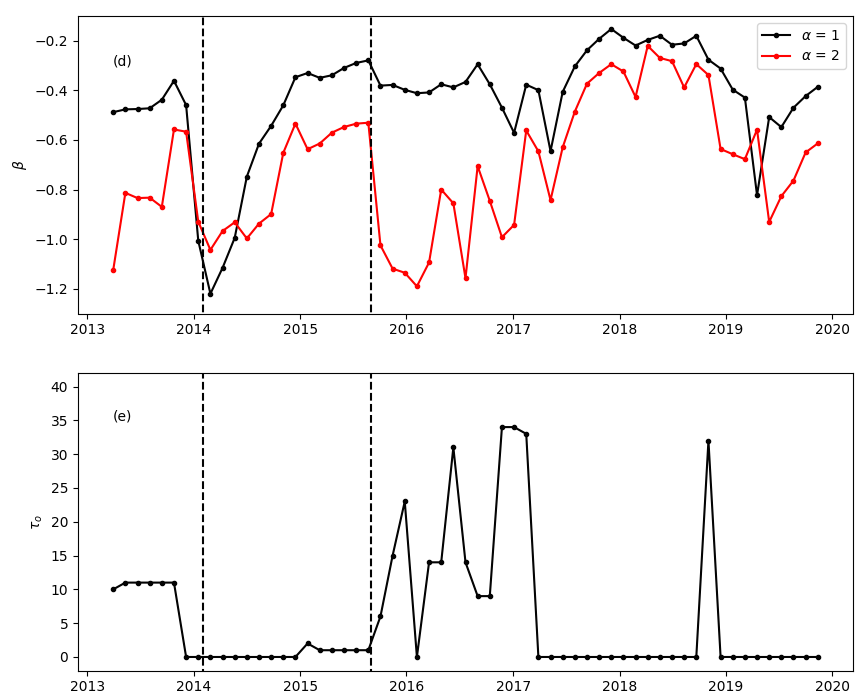}
\end{center}
\caption{\small {Running windows analysis. From top to bottom:
the dynamics of the moments of the distribution (a,b,c).
The values of the $\beta$ exponent for characterizing
the autocorrelation functions of powers of the returns (d)
and the leverage effect (e), described using the characteristic time 
$\tau_0$. Note that $\tau_0=0$ corresponds to data which do not 
present the leverage effect.
The analysis is obtained using overlapping windows of 1600 data points, 
moving forward by 200 datapoints.
The first dashed vertical line represents the
Mt. Gox incident (2/2014) and the second one the unsuccessful 
fork attempt of Bitcoin into Bitcoin XT (8/2015).
}
}
\label{fig_windows}
\end{figure}

By looking at the behaviour of the moments of the distribution
(see Fig. \ref{fig_windows}) we can note that the
volatility decreases monotonically until close to 2/2016, 
when it starts to grow.
Around the same period there is a drop towards
strong negative values of the skewness and, conversely,
a peak in the value of the kurtosis.

We estimate also the values of the $\beta$ exponents for characterizing the 
autocorrelation functions of powers of the returns
and  the characteristic time $\tau_0$ for 
describing the presence of the leverage effect (see Fig. \ref{fig_windows}).
By looking at these quantities it is possible to detect two clear
critical moments in the time evolution
of the Bitcoin market: the early 2014,
which presents an abrupt rise in the leverage effect,
which reaches a characteristic time of more than 7 days,
followed by the disappearance of the same effect.
In the same period a strong fall in the exponents of the non-linear
correlations is recorded.
This period follows the Mt. Gox incident when,
in February 2014, this Bitcoin exchange suspended trading and closed its service \cite{MtGox}.

The second critical moment is in the late 2015,
and it is characterized by high values
of the  $\beta$ exponent for $\alpha=1$  
and the reappearance of the leverage effect ($\tau_0>0$).
It can be associated with
the unsuccessful fork attempt of Bitcoin into Bitcoin XT
in August 2015.
In between these two important events, 
the Bitcoin history is characterized by relative small Kurtosis and Skewness.

Over the months following the early 2017 Kurtosis, Skewness
and $\tau_0$ approach values close to 0. In this period, the values of the $\beta$ exponents
are comparable to the ones reported in \cite{power2}.

\section{Discussion}

In this paper, we explore some stylized facts of the Bitcoin market from 2013 to 2020, using the BTC-USD exchange rate time series with a time lag of 5 hours.

We detect heavy tails in the distribution of returns, which can be characterized by a power-law. 
The tail index of the cumulated distribution is close to 3,  a value comparable to the one displayed by foreign exchange markets and stock markets \cite{guillame,gabaix}. It is interesting to note that, using 
higher frequencies data,
Bitcoin returns distributions with heavier tails ($k<2.5$)  have been recorded \cite{begusic}. 
Our result suggests that, at larger temporal scales, the expected exponent close to 3 
is recovered. 

The analysis of the behaviour of the absolute and squared returns shows important positive autocorrelations, which decay slowly as a function of the time lag, following a power-law with the exponents $0.31$, for the absolute returns, and $0.72$ for the squared returns. These results are 
homologous with the behaviour found in the time series of exchange rates of fiat money. 
By looking at the linear autocorrelations we have observed that even if they are very close to be always negligible, as expected for the exchange rates of fiat money, we can detect some values 
above the significance level. This fact is consonant with some previous studies \cite{bariviera}, in particular with the work of Urquhart \cite{Urquhart2016} which shows some antipersistent behaviors dependent on the sampling period.
More marked differences can be found in the measurement of the volatility-return correlation, which shows a negative correlation (leverage effect) with a decay time slightly longer than 2 days.
Bitcoin shows a leverage effect analogous to that of stocks and stock indices, 
in contrast with the negligible importance that traditional studies give to this effect in exchange rates. 
Our findings can be contrasted with previous results for Bitcoin time series, which claimed to have found
no leverage effect \cite{absence} or an inverted leverage effect \cite{inverted}, and
compared to a recent analysis of the conditional volatility, based on financial econometrics models, 
which suggested the possible existence of the effect  \cite{presence}.
Important differences are also present in the extreme high value of the volatility, with a difference of one order of magnitude in relation to other fiat money, and the presence of a gain/loss asymmetry, a property usually present in stock prices and stock index values, but not in exchange rates, where there is a higher symmetry in up/down movements.

Bitcoin market is characterized by some very peculiar idiosyncrasies. It seems that there are no macroeconomic fundamentals for digital currencies, nor their values can be derived from an underlying asset or benchmark. The absence of macro-financial indicators should generate more sensitivity on information flows in market and affect the supply and demand interaction. Moreover, there is a clear asymmetry between a demand, strongly influenced by speculative interests, and its rigid supply. The presence of inefficiency in the form of very strong arbitrage opportunity can be easily tracked looking at the important spread among the different platforms used for buying and selling this cryptocurrency. 
Anyway, all these elements seem to be marginal in the definition of the structural statistical properties of this virtual financial asset, which result to be analogous to general individual stocks or indices.
This fact suggests conjecturing about what determines the 
statistical properties of the considered stylized facts for a general financial asset.
As these properties do not change in the case of the Bitcoin, they
should be related more to the market and the social component shaping the price, and much less to macroeconomic aspects, or connections with underlying assets or benchmarks.

If the comparison is restricted to fiat money exchange rates time series, we can outline some clear differences in the linear autocorrelation and, most importantly, in the presence of the leverage effect, gain/loss asymmetry and the extreme high value of the volatility.
Foreign exchange rates, with their high liquidity, can be considered as a market which better supports the efficient market hypothesis. In the particular case of the Bitcoin market, the evident detection of positive autocorrelations in the absolute and squared returns and, most importantly, the presence of a clear leverage effect which can be interpreted as a sign of long-range dependence, provide evidence against the efficient market hypothesis.\\

The second part of our analysis explores the dynamics of correlations all along the evolution of the Bitcoin time series, by partitioning it 
with sliding windows of 1600 data points, which correspond to less than one year. 
We look at the behaviour of the moments of the distribution, the character of the slow decay of the absolute and squared returns autocorrelation functions (using the $\beta$ exponent) and the leverage effect (using the characteristic time $\tau_0$).
These indicators of specific statistical characters of the time series behave particularly well as empirical signals for monitoring the shifts in the evolution of the Bitcoin market. 
In fact, their changes are clearly correlated to two important critical moments of the Bitcoin dynamics: the Mt. Gox incident, 
and the unsuccessful fork attempt of Bitcoin into Bitcoin XT in August 2015.

In the interval between these two important events, the Bitcoin history is characterized by relative small kurtosis and skewness, with absence of the leverage effect and a strong fall in the exponents of the absolute return correlations.
These features generally correspond to a stochastic process with weaker memory signatures and closer to Gaussianity.
In contrast, outside this region, deviation from Gaussianity (higher skewness and kurtosis) and correlations (leverage and a slower decay in the absolute return correlation) become more relevant. We can note how Mt. Gox incident was a huge strike to the Bitcoin's credibility and reputation, which affected its price generating a shock in a moment of euphoria, which led to a stable decreasing trend. An inversion to this tendency appeared in the late 2015, with a new rise associated with an increasing popular attention. 
Following the work of Gerlach {\it et al.}, this interval corresponds to the correction regime which follows the second Bitcoin's long bubble, and the neutral period before the rise of the third long bubble \cite{Gerlach2019}.
From these considerations, we can suppose that periods of greater interest associated to speculative activities generate stronger deviation from Gaussianity and more relevant memory effects.

Similarly, over the months following the early 2017 
the absence of the leverage effect and a small 
skewness is registered. This behaviour occurs with
$\beta$ exponents values comparable to the ones of mature  
markets. These facts suggest that the Bitcoin was approaching
a maturity regime, as already pointed out by other works \cite{drozdz}.
Actually, in that period, important events which can be correlated 
with the maturation of the Bitcoin market happened:  
the breaking of the psychological $1000$ USD barrier 
and the announcement and launch of Bitcoin future contracts. 

From the results of our running windows analysis we can state that the considered statistical
indicators behave particularly well for monitoring the evolution of the Bitcoin time series and for 
characterizing the regime switches in the Bitcoin market. 
For this reason, a future analysis based on these indicators  applied to other cryptocurrencies
could be able to characterize possible spill-over effects in the cryptocurrency market. 
This is a theme of particular interest, as already pointed out by
Katsiampa {\it et al.} \cite{Katsiampa}, who suggested that there are two-way shock propagation effects between Bitcoin and Ether and Bitcoin and Litecoin. In the first case, past Bitcoin shock news has a positive effect on Ether current conditional volatility, whereas prior Ether shocks have a negative impact on Bitcoin current volatility. As far as Bitcoin and Litecoin are concerned, the lagging shocks of one digital currency adversely impact the present conditional volatility of the other. 


\begin{acknowledgements}
F.N.M.S.F. acknowledges the  PIBIC program of Universidade Federal do Rio de Janeiro for partial financial support. 
M.A.B. is grateful to Fapesp (Grant 2018/22562-4) and CNPQ (Grant 303986/2017-4 and 428433/2018-9) for their support.
\end{acknowledgements}





\end{document}